\documentclass[12pt]{article}
\usepackage{amssymb}
\def\be{\begin{equation}}
\def\ee{\end{equation}}
\def\bes{\begin{eqnarray}}
\def\ees{\end{eqnarray}}
\def\p{\partial}

\def\ket#1{|#1\rangle}
\def\half{{1\over2}}

\catcode`\@=11
\def\@citex[#1]#2{%
\if@filesw \immediate \write \@auxout {\string \citation {#2}}\fi
\@tempcntb\m@ne \let\@h@ld\relax \def\@citea{}%
\@cite{%
  \@for \@citeb:=#2\do {%
    \@ifundefined {b@\@citeb}%
      {\@h@ld\@citea\@tempcntb\m@ne{\bf ?}%
      \@warning {Citation `\@citeb ' on page \thepage \space undefined}}%
      {\@tempcnta\@tempcntb \advance\@tempcnta\@ne%
      \@tempcntb\number\csname b@\@citeb \endcsname \relax%
      \ifnum\@tempcnta=\@tempcntb %Number follows previous--hold on to it
        \ifx\@h@ld\relax%
          \edef \@h@ld{\@citea\csname b@\@citeb\endcsname}%
        \else%
          \edef\@h@ld{\ifmmode{-}\else--\fi\csname b@\@citeb\endcsname}%
        \fi%
      \else%   %  non-successor--dump what's held and do this one
        \@h@ld\@citea\csname b@\@citeb \endcsname%
        \let\@h@ld\relax%
      \fi}%
    \def\@citea{,\penalty\@highpenalty\,}%
  }\@h@ld
}{#1}}

\def\@citeb#1#2{{[#1]\if@tempswa , #2\fi}}
\def\@citeu#1#2{{$^{#1}$\if@tempswa , #2\fi }}
\def\@citep#1#2{{#1\if@tempswa , #2\fi}}

\def\bcites{         % cite with []'s
        \catcode`\@=11
        \let\@cite=\@citeb
        \catcode`\@=12
}

\def\upcites{         % cite with exponents
        \catcode`\@=11
        \let\@cite=\@citeu
        \catcode`\@=12
}

\def\plaincites{      % cite without brackets
        \catcode`\@=11
        \let\@cite=\@citep
        \catcode`\@=12
}

\newcount\hour
\newcount\minute
\newtoks\amorpm
\hour=\time\divide\hour by 60
\minute=\time{\multiply\hour by 60 \global\advance\minute by-\hour}
\edef\standardtime{{\ifnum\hour<12 \global\amorpm={am}%
        \else\global\amorpm={pm}\advance\hour by-12 \fi
        \ifnum\hour=0 \hour=12 \fi
        \number\hour:\ifnum\minute<10 0\fi\number\minute\the\amorpm}}
\edef\militarytime{\number\hour:\ifnum\minute<10 0\fi\number\minute}

\def\draftlabel#1{{\@bsphack\if@filesw {\let\thepage\relax
   \xdef\@gtempa{\write\@auxout{\string
      \newlabel{#1}{{\@currentlabel}{\thepage}}}}}\@gtempa
   \if@nobreak \ifvmode\nobreak\fi\fi\fi\@esphack}
        \gdef\@eqnlabel{#1}}
\def\@eqnlabel{}
\def\@vacuum{}
\def\marginnote#1{}
\def\draftmarginnote#1{\marginpar{\raggedright\scriptsize\tt#1}}
\def\draft{
        \pagestyle{plain}
        \overfullrule=2pt
        \oddsidemargin -.5truein
        \def\@oddhead{\sl \phantom{\today\quad\militarytime} \hfil
        \smash{\Large\sl DRAFT} \hfil \today\quad\militarytime}
        \let\@evenhead\@oddhead
        \let\label=\draftlabel
        \let\marginnote=\draftmarginnote
        \def\ps@empty{\let\@mkboth\@gobbletwo
        \def\@oddfoot{\hfil \smash{\Large\sl DRAFT} \hfil}
        \let\@evenfoot\@oddhead}
        \def\@eqnnum{(\theequation)\rlap{\kern\marginparsep\tt\@eqnlabel}%
        \global\let\@eqnlabel\@vacuum}  }
\begin{document}

%\hfill CERN-TH/2001-310

\hfill UTHET-02-0102

%\hfill {\tt hep-th/0111085} 
\vspace{-0.2cm}

\begin{center}
\Large
{\bf dS/CFT Correspondence in Two Dimensions\footnote{Research supported in part by the DoE under grant DE-FG05-91ER40627.}}
\normalsize

\vspace{0.8cm}

{\bf Scott Ness\footnote{ness@utk.edu} and George Siopsis\footnote{gsiopsis@utk.edu}}\\ Department of Physics
and Astronomy, \\
The University of Tennessee, Knoxville \\
TN 37996 - 1200, USA.
\end{center}

\vspace{0.8cm}
\large
\centerline{\bf Abstract}
\normalsize
\vspace{.5cm}
We discuss the quantization of a scalar particle moving in two-dimensional
de Sitter space. We construct the conformal quantum mechanical model on the
asymptotic boundary of de Sitter space in the infinite past. We obtain
explicit expressions for the generators of the conformal group and
calculate the eigenvalues of the Hamiltonian. We also show that two-point
correlators are in agreement with the Green function one obtains from
the wave equation in the bulk de Sitter space.

\newpage

%\section{Introduction}
%\vspace{.5cm}
Even though the AdS/CFT correspondence is by now well-understood
\cite{bh,balasubramanian,giddings,gubser,witten,hm,sw,bb}, a similar correspondence
for a de Sitter space has been quite a puzzle to establish. Such a development
is of considerable interest in view of recent astronomical data suggesting
that we live in a Universe of positive cosmological constant \cite{permutter}.

A concrete proposal
for a dS/CFT correspondence was recently put forth by Strominger \cite{strominger} and attracted much attention~\cite{ds}. According to this proposal, all observables
in the bulk de Sitter space are generated by data specified on its asymptotic
boundary which can be selected to be the Euclidean hypersurface $\mathcal{I}^-$
in the infinite past. The isometries of the de Sitter space are mapped onto
generators of the conformal group of the theory defined on the boundary $\mathcal{I}^-$. This CFT is hard to construct in general, but various features,
such as conformal weights and masses, are known. In three-dimensional de Sitter
space, the conformal group on the boundary is infinite and the central
charge is known~\cite{bms}. The CFT is a Liouville theory~\cite{ck}.

Here we discuss the case of two-dimensional de Sitter space. The asymptotic
boundary $\mathcal{I}^-$ is a circle, which upon a Wick rotation turns into
time. The theory on the boundary is a conformal quantum mechanical model.
We shall explicitly construct this model for the case of a scalar particle,
obtain the generators of the conformal group, calculate the eigenvalues of the
Hamiltonian and the Green functions. Our method of solution is similar to the one
discussed by de Alfaro, Fubini and Furlan (DFF)~\cite{dff}, even though our Hamiltonian differs from theirs.

%In this paper we follow the steps taken by Strominger \cite{strominger} and look at the de Sitter CFT correspondence for two dimensions.  We will constrain our work to only a small region of space denoted $\mathcal{O}^-$, which includes the planar past asymptotic region $\mathcal{I}^-$, but not $\mathcal{I}^+$.  This region can be sliced into many asymptotic spacelike surfaces.  It is known there exists quantum states on these spacelike slices.  We would like to find the quantum states and conformal generators that live on these slices for a two dimensional de Sitter space.

%When quantizing the wave equation we are posed with a normal ordering problem which can be understood after calculating the eigenvalue of the Casimir for the quantum system.  During this process we will calculate the conformal generators and show they satisfy the $SL(2,\mathbb{C})$ algebra.  We follow the work of \cite{dff} and calculate the ground state energy for the quatum states living on the asyptotic spacelike slices.

The two-dimensional de Sitter space (dS$_2$) may be parametrized as 
\be
  \frac{ds^2}{l^2}=-d\tau^2+\cosh^2\tau d\phi^2.
\ee
We will explicitly consider the case where $\phi$ spans the real axis (zero temperature limit) and then comment on what changes need to be made to
turn $\phi$ periodic (finite temperature).

Consider a scalar field $\Phi$ of mass $m$. It obeys the wave equation
in de Sitter space
\be\label{eqwav}
  l^2 \nabla^2\Phi=-\frac{1}{\cosh\tau}\p_\tau\left(\cosh\tau\p_\tau\Phi\right)+
  \frac{1}{\cosh^2\tau}\p^2_\phi\Phi=m^2l^2\Phi.
\ee
The operator $\nabla^2$ is the Casimir operator in the $SL(2,{\mathbb R})$ algebra
generated by
\be\label{eqgen} L_\pm = e^{\pm i\phi} \left( i\tanh\tau {\partial\over\partial\phi} \pm {\partial\over\partial\tau}\right) \quad,\quad L_3 = -i {\partial\over\partial\phi}\ee
Indeed,
\be \ell^2 \nabla^2 = \half (L_+L_-+L_-L_+) - L_3^2\ee
Focusing on the $\mathcal{I}^-$ boundary reduces the wave equation to
\be
  \left(-\p^2_\tau+\p_\tau+4e^{2\tau}\p^2_\phi\right)\Phi=m^2l^2\Phi\label{wave}.
\ee
On this spacelike slice we see the third term is negligible and the wave equation becomes independent of $\phi$. 
The solution to the wave equation~(\ref{eqwav}) behaves asymptotically as
\be\label{eqnumu}
  \Phi \sim e^{h_\pm\tau}\hspace{1cm},\hspace{1cm}h_\pm=\half \pm\nu
\hspace{1cm},\hspace{1cm}\nu = i\mu \hspace{1cm},\hspace{1cm}\mu =\half \sqrt{4m^2l^2 -1}.
\ee
We will concentrate on the case of imaginary $\nu$ ($\mu\in\mathbb{R}$) in
which $h_- = h_+^\star$.
Introducing the coordinate $q=e^\tau$, we may write~(\ref{wave}) as
\be\label{wave2} f_k'' + 4k^2 f_k =- \frac{m^2\ell^2}{q^2} \; f_k\ee
where $\Phi (\tau,\phi) = e^{ik\phi} f_k(q)$. The solutions are
\be f_k (q) = \sqrt{2kq} \; Z_\nu (2kq)\ee
where $Z_\nu$ is a Bessel function.
They form an orthogonal set under the inner product
\be (\Phi_1,\Phi_2) = i\int_{-\infty}^\infty \frac{d\phi}{2\pi}\; e^{i(k_1-k_2)\phi}\; (f_{k_1}^\star (q)
f_{k_2}' (q) - {f_{k_1}^\star}' (q) f_{k_2} (q)).\ee
The apparent $q$-dependence disappears after we apply the wave equation and the
integral over $\phi$ leads to a $\delta$-function, as expected.
For the Euclidean choice
\be \Phi_k^E = C^E e^{ik\phi} \sqrt{2kq} \; H_\nu^{(1)} (2kq)\ee
demanding orthonormality, fixes the normalization constant to $C^E = \frac{1}{\sqrt 8}$ for
real $\nu$, where we used the Wronskian $H_\nu^{(1)} (x) {H_\nu^{(2)}}' (x)
-{H_\nu^{(1)}}' (x) H_\nu^{(2)} (x) = \frac{-4i}{\pi x}$.
Another interesting choice is
\be\label{eq5} \Phi_k^+ = C^+ e^{ik\phi} \sqrt{2kq} \; J_\nu (2kq).\ee
For an orthonormal set, we choose
\be C^+ = \frac{\sqrt{\nu}}{2}\; \frac{\Gamma(\nu)}{\Gamma({\half})}\ee
%This also ensures that $\Phi_k^+$ has unit norm.
Using the Bessel function identity
\be J_\nu(x) = \half \left(H_\nu^{(1)}(x) - e^{i\pi\nu}
H_\nu^{(1)} (-x)\right)\ee
we obtain
\be\label{eq1} \Phi_k^+ (q,\phi) = \sqrt{\frac{\nu}{2\pi}}\Gamma(\nu) \left( \Phi_k^E (q,\phi) +
e^{i\pi h_+} \Phi_k^E (-q,\phi)\right)\ee
%forming an orthonormal set.
which differs from \cite{bms} by a phase.
The Green function for the modes $\Phi_k^+$ can be obtained from
\be G^+ (q,\phi; q',\phi') = \int \frac{dk}{k} \Phi_k^+ (q,-\phi)
\Phi_k^+ (q',\phi').\ee
After some algebra, we arrive at
%is easily obtained from the wave equation~(\ref{eqwav})
\be G^+(q,\phi; q',\phi') = (C^+)^2\frac{\Gamma (h_+)}{\Gamma(h_++\half)\Gamma(\half)}\; (4P)^{-h_+}\; F(h_+,h_+; 2h_+; -1/P)\ee
where $P$ measures the distance between the two arguments of the Green
function in the three-dimensional Minkowski space in which dS$_2$ is embedded,
\be P = \frac{4(q-q')^2 + (\phi-\phi')^2}{16qq'}.\ee
When $P\to \infty$, the Green function behaves as
\be G^+(q,\phi; q',\phi') \sim (C^+)^2\frac{\Gamma (h_+)}{\Gamma(h_++\half)\Gamma(\half)}\; (4P)^{-h_+}. \ee
For the Euclidean modes, the Green function can be found in terms of the
Green function obtained above using~(\ref{eq1})~\cite{strominger}. Near the
boundary, we deduce the two-point function of the dual conformal
theory,
\be\label{eqgp} \langle {\cal O}^\dagger (\phi) {\cal O}(\phi') \rangle = 2\sqrt\pi\frac{\Gamma (h_+)}{\Gamma (\nu)}\; (\phi-\phi')^{-2h_+}\ee
%which is the two-point function of the dual conformal theory.
%\be \langle {\cal O} {\cal O} \rangle = \ee
%The de Sitter invariant Green function was calculated in D dimensions by Strominger $\cite{strominger}$.  Near the boundary the Green function is given by
%\be
%  \lim_{\tau,\tau'\rightarrow\infty}G[\tau,\phi,\tau',\phi']=\frac{16^{h_+}\Gamma(h_--h_+)}{\Gamma(h_-)\Gamma(1-h_-)}\frac{e^{h_+(\tau+\tau')}}{(\phi-\phi')^{2h_+}}+(h_+ \leftrightarrow h_-).
%\ee
If $\phi$ is a periodic coordinate with period $2\pi$, then $k$ takes on
discrete values and the integral defining the Green function turns into a sum.
The form of the two-point function may be deduced from the periodicity condition and the singularity, which uniquely determine it. We obtain
\be\label{eqgpf} \langle {\cal O}^\dagger (\phi) {\cal O}(\phi') \rangle = 2\sqrt\pi\frac{\Gamma (h_+)}{\Gamma (\nu)}\; \left( 2\sin\frac{\phi-\phi'}{2}\right)^{-2h_+}\ee
%--------- Classical picture -------
The boundary conformal theory can be defined in terms of the
coordinates $q=e^\tau,\ \phi$ and their conjugate momenta $p=e^{-\tau}\p_\tau,\ H$, respectively.  Instead of the
$SL(2,{\mathbb R})$ generators~(\ref{eqgen}), it is more convenient to
introduce the generators
%We can also define two more quantities D and K such that
\bes\label{eq2}
  D &=& i\left(\phi\p_\phi+\p_\tau\right)\nonumber \\
  K &=& -\phi^2\p_\phi+2i\phi\p_\tau+4e^{2\tau}\p_\phi.
\ees
in the vicinity of the boundary $\mathcal{I}^-$.
The three operators H,D and K satisfy the $SL(2,\mathbb{R})$ algebra 
\be
  [H,K]=2iD,\hspace{1cm}[H,D]=iH,\hspace{1cm}[K,D]=-iK.
\ee
The wave equation is mapped onto the constraint
\be
  4q^2H^2-q^2p^2= g \quad,\quad g= m^2\ell^2+A \label{cc}
\ee
where we have allowed for the possibility of a normal ordering ambiguity by
introducing the constant $A$.
Classically, we can solve the constraint~(\ref{cc}) and express the conformal Hamiltonian as
%and see the dependence on a conformal Hamiltonian
\be
  H=\half\sqrt{p^2+\frac{g}{q^2}}\label{H}.
\ee
The constraint thus reduces the system to a quantum mechanical model with
a single degree of freedom $q$, with $\phi$ playing the role of imaginary time.
Notice that our Hamiltonian differs from
the one considered by DFF~\cite{dff} (theirs is the square of ours), yet the results
are similar.

In order to quantize this system, we promote $(q,p)$ to operators $(Q,P)$ and
impose the standard commutation relation
\be
  [Q,P]=i
\ee
% we must treat our coordinates $q$ and $\phi$ as operators and impose the commutation relations
%\be
%  [Q,P]=i,\ \ [\phi,H]=i, 
%\ee
The coordinate $\phi$ is identified with time after a Wick rotation.
%where $P$ and $H$ are conjugate to $Q$ and $\phi$ respectively.  The wave equation can now be written as the constraint equation 
%
%\be
%  4Q^2H^2=Q^2P^2+g.\label{qc}
%\ee
%This constraint reduces the two degrees of freedom down to one on the boundary.  Quantum mechanically the constraint is not equal to equation ($\ref{cc}$).  This comes from the fact that when quantizing the system we introduce a normal ordering ambiguity.  Therefore equation ($\ref{qc}$) can be equal to equation ($\ref{cc}$) up to some normal ordering constant which will be found later.
%In terms of of the new operators, 
$D$ and $K$ (eq.~(\ref{eq2})) become the dilation and special conformal operators, respectively, given by
\bes
  D &=& i\left(\phi H+\half(QP+PQ)\right) \nonumber \\
  K &=& -\phi^2 H+i\phi(QP+PQ)+2(Q^2H+HQ^2).
\ees
%On the $\mathcal{I}^-$ boundary we have a one dimensional circle parametrized by $\phi$.  We can treat $\phi$ as time on the boundary quantum mechanical system.  
Following the procedure in~\cite{dff}, it is more convenient to start with the
values of the $SL(2,\mathbb{R})$ generators at $\phi=0$. Let $Q_0=Q(0)$ and $P_0=P(0)$.  The three generators reduce to
\bes\label{eqgen1}
  H &=& \half\sqrt{P_0^2 + \frac{g}{Q_0^2}} \nonumber \\
  D &=& {i\over 2}(Q_0P_0+P_0Q_0) \nonumber \\
  K &=& 2(Q_0^2H+HQ_0^2).
\ees
It can be checked explicitly that they obey the $SL(2,\mathbb{R})$ algebra
\be
  [H,K]=-2iD,\hspace{1cm}[H,D]=-iH,\hspace{1cm}[K,D]=iK.
\ee
The Casimir operator of this algebra is given by
\be\label{eqcas}
  N=\half(HK+KH)-D^2.
\ee
This commutes with all operators and its numerical value should equal $m^2\ell^2$ to match the wave equation. To calculate $N$, we first bring it into
the form
\be N = [H,\;[Q_0^2,H]]+g+{3\over 4}\ee
after some straightforward operator manipulations.
The remaining double commutator is independent of $g$ and can therefore be
easily calculated by setting $g=0$ in $H$. It may also be deduced from the
corresponding Poisson brackets. The final result is
\be N = g - \frac 1 4\ee
By comparing with the wave equation~(\ref{wave}), we deduce that the normal ordering
constant is $A=1/4$~({\em cf.}~(\ref{cc})) and
\be g = m^2\ell^2 + \frac 1 4\ee
We follow the procedure of \cite{dff} to construct the states in the Hilbert space. Define new operators R, S:
\bes
  R &=& \half\left({1\over a}K+aH\right), \nonumber\\
  S &=& \half\left({1\over a}K-aH\right).
\ees
where $a$ is a constant with dimension of length.
We will let $a=1$ for simplicity.
$R$ plays the role of a Hamiltonian for the quantum mechanical system on $\mathcal{I}^-$ whose eigenstates span the Hilbert space.
By including the dilation operator, $R,S$ and $D$ form the closed algebra $O_{2,1}$,
\be
  \left[D,R\right]=iS\hspace{1cm}\left[S,R\right]=-iD\hspace{1cm}\left[S,D\right]=-iR.
\ee
Next, we define raising and lowering operators
\be
  L_\pm =S\mp iD.
\ee
These new operators, along with $R$ form the closed algebra $SL(2,\mathbb{R})$,
\be
  [R,L_\pm ]=\pm L_\pm\hspace{1cm}\left[L_+,L_-\right]=-2R.
\ee
The Casimir operator for this algebra (\ref{eqcas}) can be written
in terms of $R,L_+$ and $L_-$ as
\be
  N=\half(HK+KH)-D^2\;=R^2-R-L_+L_-
\ee
The ground state energy can be found by operating on the ground state of $R$,
\be
  R\ket{0}=r_0\ket{0},
\ee
\be
  N\ket{0}=r_0(r_0-1)\ket{0}=m^2l^2\ket{0}.
\ee
Thus, the ground state energy is
\be
  r_0=\half(1\pm\sqrt{1-4m^2l^2}).
\ee
Therefore, $r_0 = h_\pm$ (eq.~(\ref{eqnumu})). Without loss of generality,
we choose $r_0 = h_+$. The other choice ($r_0=h_-=h_+^\star$) leads to
a dual Hilbert space related to the one we are about to study via complex
conjugation. It should be noted that inner products ought to be defined
in terms of both Hilbert spaces, leading to normalization conditions such as
(\ref{eqnn}).

By successively acting with the raising operators, we deduce the eigenvalues
of $R$,
\be
  R\psi_n(q) = (r_0+n)\psi_n(q)
\ee
In the position representation, we obtain
the Schr\"odinger equation
\be\label{eqsch}
  \left((1+2q^2)\sqrt{-\frac{d^2}{dq^2}+\frac{g}{q^2}}+ 2\sqrt{-\frac{d^2}{dq^2}+\frac{g}{q^2}}\; q^2\right)\psi_n(q) = \sqrt{2}(r_0+n)\psi_n(q).
\ee
%The behavior of the wave function depends on g when $x\rightarrow 0$.  We find the lowest eigenfunction behaves as
%\be
%  \psi_0(q)\sim q^{2r_0-\half}.
%\ee
To solve the Schr\"odinger equation, it is convenient to introduce the
transition matrix $\mathcal{T}_{nk}$ between the two bases $\Phi_k^+$~(\ref{eq5}) and $\psi_n$, at time $\phi =0$,
\be \psi_n (q) = \int_0^\infty dk \mathcal{T}_{nk}^\star \Phi_k^+ (q),\ee
where we impose orthonormality,
\be\label{eqnn} \int_0^\infty dq\;\psi_n(q)\psi_{n'}(q)=\delta_{nn'}.\ee
%\be \mathcal{T}_{nk} = \int_0^\infty dq \psi_n^\star (q) \Phi_k^+ (q)\ee
The action of the three $SL(2,\mathbb{R})$ generators on $\Phi_k^+$ translates
to an action on the matrix elements $\mathcal{T}_{nk}$,
\be\label{eq50}
\hat H \mathcal{T}_{nk} = k \mathcal{T}_{nk}\quad,\quad \hat D \mathcal{T}_{nk} = -i\left( k\frac{d}{dk} + \half\right) \mathcal{T}_{nk}\quad,\quad \hat K \mathcal{T}_{nk} = \left( -k \frac{d^2}{dk^2} - \frac{d}{dk} + \frac{\nu^2}{k}\right) \mathcal{T}_{nk},\ee
respectively.
In this representation, the Schr\"odinger equation~(\ref{eqsch}) becomes
\be\half (\hat H + \hat K) \mathcal{T}_{nk} = (h_++n) \mathcal{T}_{nk}\ee
which can be solved. The solution is given in terms of Laguerre polynomials~\cite{dff},
\be \mathcal{T}_{nk} = \frac{2^{h_+}}{C^+} \sqrt{\frac{n!}{\Gamma (n+2h_+)}} k^\nu e^{-k}
L_n^{2\nu} (2k)\ee
where the normalization has been fixed as in~\cite{dff} with the exception of
the factor $C^+$ which is due to the different normalization condition we imposed on the wavefunctions $\Phi_k^+$. It leads to a unit norm for the wavefunction $\psi_n$.
%The solution to the Schr\"odinger equation may be obtained from the integral
%\be \psi_n (q) = \int_0^\infty dk \mathcal{T}_{nk}^\star \Phi_k (q)\ee
It is convenient to
introduce the Laplace transform of $\mathcal{T}_{nk}$ with respect to the energy $k$,
\be \widetilde{\mathcal{T}}_n (\phi) = 2^{-h_+} \int_0^\infty dk e^{-k\phi} \; k^\nu\; \mathcal{T}_{nk}\ee
given in terms of the conjugate time variable $\phi$. The additional energy
factor $k^\nu$ ensures that the action of the $SL(2,\mathbb{R})$ generators
is as expected.
Explicitly,
\be \widetilde{\mathcal{T}}_n (\phi) = \frac{(-)^n}{(C^+)^2} \sqrt{\frac{\Gamma(n+2h_+)}{n!}}
\; \left(\frac{1-\phi}{1+\phi}\right)^{n+h_+}\; (1-\phi^2)^{-h_+}\ee
and the $SL(2,\mathbb{R})$ generators act as ({\em cf.}~eq.~(\ref{eq50}))
\be \hat H \widetilde{\mathcal{T}}_n = -\frac{d}{d\phi} \widetilde{\mathcal{T}}_n\quad,\quad \hat D \widetilde{\mathcal{T}}_n = i\left( \phi\frac{d}{d\phi} + h_+\right) \widetilde{\mathcal{T}}_n\quad,\quad \hat K \widetilde{\mathcal{T}}_n = \left( \phi^2 \frac{d}{d\phi} + 2h_+\phi\right) \widetilde{\mathcal{T}}_n.\ee
%normalized so that
%\be \sum_{n=0}^\infty\; \mathcal{T}_{nk}^\star \mathcal{T}_{nk'} = \delta (k-k')\ee
The two-point function is given by
\be G(\phi,\phi') = \sum_{n=0}^\infty \widetilde{\mathcal{T}}_n (\phi) \widetilde{\mathcal{T}}_n (-\phi')\ee
Notice that we there is no complex conjugation on $\widetilde{\mathcal{T}}_n (-\phi')$, because it ought to be the complex conjugate of the dual of $\widetilde{\mathcal{T}}_n$ which is itself related to $\widetilde{\mathcal{T}}_n$ by complex
conjugation ({\em cf.}~also the inner product~(\ref{eqnn})). A short calculation reveals
\be G(\phi,\phi') = \frac{\Gamma (2h_+)}{2^{2h_+}(C^+)^2}\; (\phi-\phi')^{-2h_+}
= 2\sqrt\pi\frac{\Gamma(h_+)}{\Gamma (\nu)}\; (\phi-\phi')^{-2h_+}\ee
in agreement with the de Sitter space result (\ref{eqgp}). Higher-order correlators
may also be obtained by using the standard methods developed in~\cite{dff}.

If we demand periodicity of $\phi$ ($\phi \equiv \phi + 2\pi$), the above calculations become cumbersome. The energy $k$ takes on discrete values and
the differential equations determining the transition matrix $\mathcal{T}_{nk}$
turn into difference equations. Nevertheless, correlators may be deduced without explicit calculations, because they are uniquely determined by the
periodicity requirements and their singularities.

%Next, we turn to the case where the temperature is finite.
To summarize, we have discussed the conformal quantum mechanical model which
resides on the boundary $\mathcal{I}^-$
of two-dimensional de Sitter space in the infinite past. We calculated the eigenvalues and corresponding eigenfunctions of the
Hamiltonian and deduced correlation functions. We showed that the Green functions agree with the propagators one obtains from the de Sitter space wave equation.
Most of our results are similar
to those of ref.~\cite{dff}, even though our Hamiltonian is different and the
Schr\"odinger equation~(\ref{eqsch}) cannot be solved explicitly. What saves the day is the
$SL(2,\mathbb{R})$ symmetry which determines much of the structure of the
conformal theory as was shown in \cite{dff}. It would be interesting to
extend our results to de Sitter spaces of dimension higher than two and shed some light on
the behavior of the conformal field theories on the boundary.

\end{document}